\documentclass[a4paper,10pt]{article}
\usepackage{mathrsfs}
\usepackage{amsfonts}

\usepackage{epsf}
\usepackage{graphicx}    

\newcommand{\be}[1]{\begin{equation}\label{#1}}
\newcommand{\ee}{\end{equation}}
\newcommand{\bea}{\begin{eqnarray}}
\newcommand{\eea}{\end{eqnarray}}

\def\gsim{ \lower .75ex \hbox{$\sim$} \llap{\raise .27ex \hbox{$>$}} }
\def\lsim{ \lower .75ex \hbox{$\sim$} \llap{\raise .27ex \hbox{$<$}} }

\renewcommand{\markright}{\markright{\thepage}}

\begin{document}

\begin{titlepage}

\vspace{5mm}

\begin{center}

{\Large \bf Cyclic Universe with Quintom matter in Loop Quantum
Cosmology}

\vspace{10mm}

{\large Hua-Hui Xiong$^1$\footnote{xionghh@mail.ihep.ac.cn},
Taotao Qiu$^1$\footnote{qiutt@mail.ihep.ac.cn}, Yi-Fu
Cai$^1$\footnote{caiyf@mail.ihep.ac.cn},  Xinmin
Zhang$^{1,2}$\footnote{xmzhang@mail.ihep.ac.cn}}

\vspace{5mm}{\em \small $^{1}$ Institute of High Energy Physics,
Chinese Academy of Sciences, P.O. Box 918-4, Beijing 100049, P. R. China\\
$^{2}$ Theoretical Physics Center for Science Facilities (TPCSF),
CAS, P. R. China}

\end{center}

\vspace{5mm}

\begin{abstract}

In this paper, we study the possibility of model building of
cyclic universe with Quintom matter in the framework of Loop
Quantum Cosmology. After a general demonstration, we provide two
examples, one with double-fluid and another double-scalar field,
to show how such a scenario is obtained. Analytical and numerical
calculations are both presented in the paper.

\end{abstract}

\end{titlepage}

\newpage

\setcounter{page}{2}

\section{Introduction}

Quintom model of dark energy \cite{Feng:2004ad} has a salient
feature that its equation-of-state (EoS) crosses smoothly  over the
cosmological constant barrier $w = -1$. Phenomenologically, this
kind of model is mildly favored by the current observational data
fitting \cite{Zhao:2006qg}, but theoretically, the model building of
Quintom dark energy is a challenge due to the No-Go theorem
\cite{Xia:2007km} (also see Refs. \cite{Feng:2004ad, Vikman:2004dc,
Hu:2004kh, Caldwell:2005ai, Zhao:2005vj, Kunz:2006wc}). This No-Go
theorem forbids a traditional scalar field model with Lagrangian of
the general form ${\cal L} = {\cal
L}(\phi,\partial_{\mu}\phi\partial^{\mu}\phi)$ from having its EoS
across the cosmological constant boundary. Consequently, in order to
realize a viable Quintom model in the framework of Einstein's
gravity theory, we need to add extra degrees of freedom to the
conventional single field theory. The simplest Quintom model
involves two scalars with one being Quintessence-like and another
Phantom-like \cite{Feng:2004ad, Guo:2004fq}. This model has been
studied in detail in \cite{Quintom_tf, Cai:2006dm}. In recent years
there have been a lot of theoretical studies of Quintom-like models.
For example, motivated from string theory, the authors of Ref.
\cite{Cai:2007gs} realized a Quintom scenario by considering the
non-perturbative effects of a DBI action. Moreover, there are models
which involve higher derivative terms for a single scalar field
\cite{Li:2005fm}, models with vector field
\cite{ArmendarizPicon:2004pm}, making use of an extended theory of
gravity \cite{Cai:2005ie}, non-local string field theory
\cite{Aref'eva:2005fu}, and others (see e.g. \cite{Quintom_1,
Onemli:2004mb, Quintom_others}). Due to this property, Quintom
scenario has some interesting applications in cosmology. For
example, a recent study has shown that a universe dominated by
Quintom matter can provide a bouncing cosmology which allows us to
avoid the problem of the initial singularity. Quintom Bounce,
proposed in \cite{Cai:2007qw} with its perturbation developed in
\cite{Cai:2007zv}, takes place in the early time of the universe
when the corresponding energy scale becomes very high. Therefore it
may be accompanied by quantum effects of gravity.

One possible approach to investigating quantum gravity effects in
the early universe is Loop Quantum Cosmology (LQC) \cite{LQC}
which is based on the theory of Loop Quantum Gravity (LQG)
\cite{LQG}. Beyond the usual knowledge of spacetime, LQC predicts
that the underlying geometry is discrete and the big bang
singularity is replaced by a big bounce when the energy density of
the universe approaches the Planck scale \cite{Ashtekar-old,
Ashtekar-improved}. Therefore, the quantum geometry plays a
significant role on determining the evolution of the universe in
the early time. Interestingly, one notes that effects of loop
quantum gravity also affect the evolution of Quintom universe in
late time \cite{Zhang:2007an}, since the energy density of phantom
component usually increases during the expansion of the universe.
In this paper we extend the idea of Quintom Bounce by considering
a universe filled with Quintom matter in the frame of LQC. We
interestingly find that the universe possesses an exactly cyclic
evolution such that its scale factor undergoes contracting and
expanding periodically.

In the context of LQC there exist two distinct quantum geometry
modifications, namely the inverse volume modification
\cite{Quantization-ambiguity, inversevolume} and the quadratic
density correction \cite{Ashtekar-improved}. Based on the inverse
volume modification authors of Ref. \cite{oscillating-universe}
obtained an cosmological solution with oscillating behavior for the
closed universe(see Ref. \cite{loop-branecyc} for cyclic universe in
flat model). The inverse volume modification appears by choosing a
large SU(2) representation for the holonomy in the matter part of
Hamiltonian constraint. As pointed out in \cite{inversevolume}, it
is natural to take the same spin representation for the
gravitational sector as for matter sector in the whole Hamiltonian
constraint. In this paper, we work with the fundamental spin
representation both for the matter sector and gravity sector. So,
the quantum geometry modification only comes from the quadratic
density correction which arises from the ``minimal area gap" in LQG.
In this paper we consider quadratic density correction of LQC for
the spatially flat universe.

This paper is organized as follows. In Sec. II, we study the
possibility of building the models of the cyclic universe.
Firstly, we will review briefly on LQC, then based on the modified
Friedmann equation from LQC we will present the general picture of
the cyclic universe, followed by two examples. The last section is
the discussion and conclusion.

\section{Cyclic Universe with Quintom Matter}

\subsection{Loop quantum cosmology}

LQG is a non-perturbative and background independent approach to
quantizing classical gravity canonically. LQC inherits the
framework developed from LQG, and so describes a quantized
isotropic universe. In LQG, the phase space of the classical
general relativity is described by SU(2) connection $A_a^i$ and
densitized triads $E_i^a$. Considering a homogeneous and isotropic
universe, such a symmetry of spacetime reduces the phase space of
infinite degrees of freedom to be finite. Therefore, in LQC the
classical phase space consists of the conjugate variables of the
connection $c$ and triad $p$, which satisfy Poisson bracket
$\left\{ c,p\right\} =\frac 13\gamma \kappa $, where $\kappa =8\pi
G$ ($G$ is the gravitational constant) and $\gamma $ is the
Barbero-Immirzi parameter which is fixed to be $\gamma \approx
0.2375$ by the black hole thermodynamics. For the model of flat
universe, the new variables obey the relations:
\begin{equation}
c=\gamma \dot{a},\quad p=a^2~,  \label{a8}
\end{equation} where $a$ is the scale
factor of Friedmann-Robertson-Walker (FRW) universe.
 In terms of
the connection and triad, the classical Hamiltonian constraint is
given by \cite{mathematical-structure}
\begin{equation}
{\cal H}_{cl}=-\frac 3{\kappa \gamma ^2}\sqrt{p}c^2+{\cal H}_M.
\label{a9}
\end{equation}

As the same as in LQG there is no operator corresponding to
connection $c$. For quantization the elementary variables are
triad and holonomies of connection along an edge which is defined
as $h_i\left( \mu \right) =\cos \left( \mu c/2\right) +2\sin
\left( \mu c/2\right) \tau _i$, where $\mu $ is the length of the
$i$th edge with respect to the fiducial metric, and $\tau _i$ is
related to Pauli matrices. The holonomies and the triads have well
defined quantum operators such that for quantization the
Hamiltonian constraint must be reformulated as the elementary
variables, i.e., the holonomies and triad. In order to intimately
mimic the quantization procedure of the full theory (LQG), the
Hamiltonian constraint for quantization takes the full form as
done in LQG \cite{mathematical-structure}, which can be reduced to
the Eq. (\ref{a9}). The Hamiltonian constraint operator can be
obtained by promoting the holonomies and triad to the
corresponding operators. After quantization, the underlying
geometry in LQC is also discrete as in the full theory, and the
quantum difference equation governs the evolution of the universe.
The quantum difference equation incorporating this discreteness
can evolve through the ``big bang point" without singularity.

Both in the improved framework of LQC and its original version,
the semiclassical state is constructed, the results show that by
evolving the semiclassical state backward in the expanding
universe on the order of Planck scale the universe is bounced into
a contracting branch \cite{Ashtekar-old, Ashtekar-improved}.
Furthermore, recently the bouncing behavior of LQC is confirmed in
a solvable model \cite{Ashtekar-solvable}. It also indicates that
the quantum feature of the universe can be well described by the
effective theory which predicted the modified Friedmann equation.
For the effective theory with the length scale larger than the
Planck length the spacetime recovers the continuum, and the
dynamical equation takes the usual differential form.

The effective Hamiltonian constraint is given by
\cite{effective-theory}
\begin{equation}
{\cal H}_{eff}=-\frac 3{\kappa \gamma ^2\bar{\mu}^2}\sqrt{p}\sin ^2\left( \bar{\mu}%
c\right) +{\cal H}_M~,  \label{a10}
\end{equation}
where $\bar{\mu}$ is the edge length of the square loop along which
the holonomies are computed. The physical area of the square loop is
fixed by the minimal area eigenvalue in LQG and is given by
$\bar{\mu}^2p=\alpha \ell
_{\mathrm{Pl}}^2$ ($\alpha $ is of order one and $\ell _{\mathrm{Pl}}=\sqrt{G\hbar }$ with $%
\hbar $ the reduced Planck constant). Here, ${\cal H}_M$ is
expressed by matter Hamiltonian which takes the same expression as
its classical form. By the Hamiltonian constraint (\ref{a10}) one
can get the Hamiltonian equation
\begin{equation}
\dot{p}=\left\{ p,{\cal H}_{eff}\right\} =-\frac{\gamma \kappa
}3\frac{\partial}{\partial c} {\cal H}_{eff}~.  \label{b1}
\end{equation}
Squaring the above equation and making use of the weakly vanishing
Hamiltonian constraint ${\cal H}_{eff}\approx 0$, the modified
Friedmann equation can be obtained as
\begin{equation}
H^2=\frac \kappa 3\rho  \left( 1-\frac{\rho}{\rho _c}\right)~,
\label{b2}
\end{equation}
where $\rho _c=\frac 3{\kappa \gamma ^2\bar{\mu}^2a^2}=\frac
3{\kappa \gamma ^2\alpha \ell _{\mathrm{Pl}}^2}$,
$\rho=\frac{{\cal H}_M}{a^3}$, and $H=\dot a/a$ is the Hubble
parameter. As analyzed in \cite{singh-bounce}, the modified
Friedmann equation predicts a nonsingular bounce when the matter
density approaches the critical value $\rho _c$.

In the context of the effective Hamiltonian, the matter
Hamiltonian equations behave as its classical ones, so the energy
conservation equation is still satisfied
\begin{equation}
\dot{\rho}+3H\left( \rho +P \right) =0~.  \label{b3}
\end{equation}

\subsection{The general picture of cyclic universe}

To start, we shall analyze in general how a cyclic universe works
with Quintom matter in the framework of LQC. From Eq. (\ref{b2}),
we can read that the Hubble parameter $H$ happens to be zero when
$\rho=\rho_c$. Furthermore, we can take the differentiations of
both sides of Eq. (\ref{b2}) with respect to the cosmic time, and
using Eq. (\ref{b3}), and then have
\begin{equation}
\dot H=-\frac{\kappa}{2}(1-\frac{2\rho}{\rho_c})(1+w)\rho~.
\end{equation}
From the expression above we can see that there is $\dot
H=\frac{\kappa}{2}(1+w)\rho_c$ when $\rho=\rho_c$. So at this
point, if $w<-1$, we have $\dot H<0$ and thus a turnaround
happens; on the other hand if $w>-1$, we have $\dot H>0$ and
correspondingly a bounce occurs.

The Quintom matter contains both quintessence- and phantom-like
components, and for the former, the energy density grows in the
contracting phase and decays in the expanding phase, while for the
latter, the case is opposite \footnote{The two components are
related with a cosmic duality \cite{Cai:2006dm}.}. So let's assume
the universe is expanding at the beginning without losing
generality. The quintessence-like component is decaying while
phantom-like component is growing and gradually dominating the
universe, making the total energy density growing and total EoS less
than $-1$. When the total energy density reaches the critical value
$\rho_c$, the turnaround happens. The universe ceases expanding and
turns to a contracting phase. In this phase, phantom-like component
shows the decaying behavior and quintessence-like component becomes
growing and finally dominating the universe which gives rise to the
possibility that its energy density arrives at the critical point
$\rho=\rho_c$ again. However, opposite to the case of expanding
phase the EoS of the universe becomes larger than $-1$ this time,
and hence a bounce takes place with the universe reentering the
expanding phase. To conclude, a Quintom universe in LQC can evolve
cyclically.

In the following section, we provide two examples of Quintom
matter which can realize a cyclic universe in the framework of
LQC.

\subsection{Example I: two-fluid Quintom matter in LQC }

At first, we consider the case of Quintom that consists of two
perfect fluid. For simplicity, we take both the two components to
have constant EoS, with one being larger than $-1$ and the other
being less than $-1$. The total EoS can cross $-1$ occasionally,
which is decided by the evolution of both components. So according
to the energy conservation equation, the total energy density of
this kind of Quintom should be $\rho=\rho_1+\rho_2=\rho_{1_0}
a^{-3(1+w_1)}+\rho_{2_0}a^{-3(1+w_2)}$, where the subscript ``1" and
``2" denote quintessence-like and phantom-like component
respectively and ``0" stands for the initial value.

We begin our study in one of the expanding phase, just as where we
are standing now. Since $a$ is expanding, $\rho_1$ is decreasing
and $\rho_2$ is increasing, and after a period of evolution the
phantom-like component will dominate the universe. When the energy
density reaches the critical value $\rho_c$, as discussed above,
the Hubble parameter will decrease to zero, and then change from
above zero to below zero, thus turnaround happens, and the
universe turns into a contracting phase. In this phase, however,
as the scale factor is contracting, $\rho_1$ will be increasing
and $\rho_2$ will be decreasing, and from the same logic, the
universe will be dominated by the quintessence-like component.
When the total energy density reaches $\rho_c$ again, Hubble
parameter will vanish and a bounce occurs, which drives the
universe into another expanding phase. Along this way, the
expanding and contracting phase will take place alternately,
giving the scenario of a cyclic universe.

In Fig. 1, we plot the solution of the numerical calculation. From
the figure, we can see that this kind of matter has an oscillating
EoS $w$ around $-1$, being a Quintom matter. Moreover, its energy
density oscillates between a small positive value and $\rho_c$.
Driven by it, the scale factor of the universe (as well as the
Hubble parameter) oscillates gracefully, showing a cyclic universe
scenario.

\begin{figure}[htbp]
\includegraphics[scale=0.8]{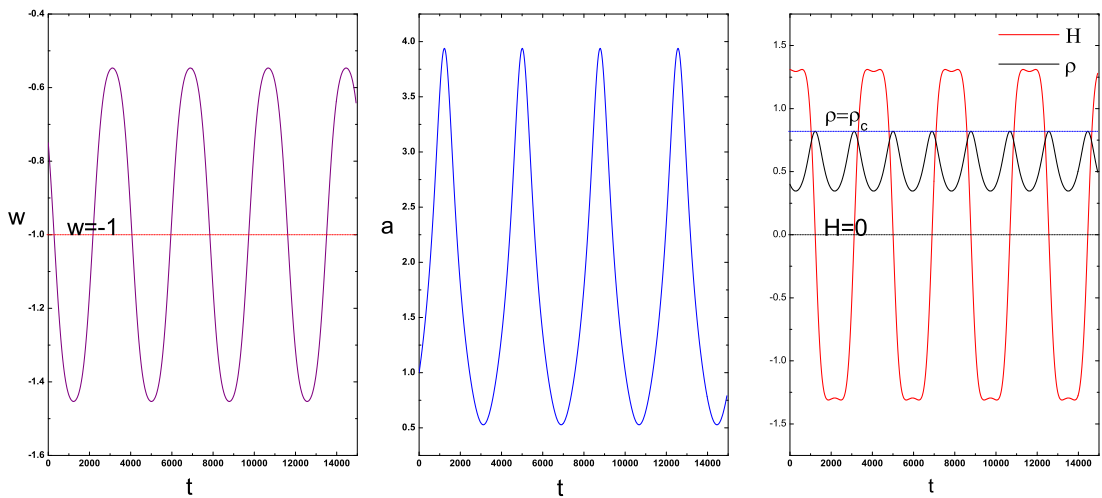}
\caption{Plots of the evolution of the EoS parameter $w$, scale
factor $a$, Hubble parameter $H$ and energy density $\rho$ in
two-fluid Quintom model. In the numerical calculation we take
$w_1=-0.5$, $w_2=-1.5$ and choose the initial values of parameters
to be $\rho_{1_0}=0.3$, $\rho_{2_0}=0.1$, $a=1.0$.}
\label{fig2:eos}
\end{figure}

\subsection{Example II: double-field Quintom matter in LQC }
We now consider the Quintom model with two scalar fields. In
spatially flat FRW cosmology, the line element is
\begin{equation}
ds^2=dt^2-a^{2}(t)(dx^2+dy^2+dz^2)~.\label{c1}
\end{equation}
We take the Lagrangian of the Quintom to be as:
\begin{eqnarray}
{\cal L} &=&
\sqrt{-g}(\frac{1}{2}\dot{\phi}^2-\frac{1}{2}\dot{\psi}^2-V(\phi,~\psi))~\nonumber
\\ &=&
\frac{1}{2}a^3\dot{\phi}^2-\frac{1}{2}a^3\dot{\psi}^2-a^3V(\phi,~\psi)~,
\end{eqnarray}
where we will choose the form of potential to be
$V(\phi,~\psi)=V_0[1+\cos(\lambda\phi\psi)]+\frac{1}{2}m_1^2\phi^2-\frac{1}{2}m_2^2\psi^2$.
In \cite{Feng:2004ad} (see also \cite{Quintom_tf}), it has already
been proven that it is the simplest Quintom model. Here, for the
cyclic universe the potential is taken to be interacting form. The
conjugate momenta are given by
\begin{eqnarray}
&\Pi_\phi=\frac{\partial L}{\partial \dot\phi}=a^3\dot\phi~,&
\nonumber\\
&\Pi_\psi=\frac{\partial L}{\partial \dot\psi}=-a^3\dot\psi~.&
\end{eqnarray}
The whole Hamiltonian can be obtained as
\begin{equation}
{\cal
H}=\frac{1}2a^{-3}\Pi_\phi^2-\frac{1}2a^{-3}\Pi_\psi^2+a^3V(\phi,~\psi)~,\label{c2}
\end{equation}
and the total energy density is
\begin{equation}
\rho=\frac{1}{2}\dot\phi^2-\frac{1}{2}\dot\psi^2+V(\phi,~\psi)~.
\end{equation}
According to the Hamiltonian Eq.(\ref{c2}) the equations of motion
for the Quintom matter are
\begin{eqnarray}
&\ddot{\phi}+3H\dot{\phi}=V_0\lambda\psi
\sin\left(\lambda\phi\psi\right)-m_1^2\phi~,&\\
&\ddot{\psi}+3H\dot{\psi}=-V_0\lambda\phi
\sin\left(\lambda\phi\psi\right)-m_2^2\psi~.&
\end{eqnarray}

In this model, there is an oscillating interaction term in the
potential, which can cause the energy density oscillating to get a
cyclic scenario. In our calculation, we found interestingly that by
appropriately choosing the parameters in the potential, we can
obtain an analytical solution. For example, if we choose $m_1=m_2=m$
and some appropriate initial conditions, we have
\begin{equation}
\phi=\phi_0\sin(mt)~,~~~~~~\psi=\psi_0\cos(mt)~,
\end{equation}
where $\phi_0$ and $\psi_0$ are constants. Using these results, we
can easily calculate the Hubble parameter $H$ with the help of
Eq.(\ref{b2}), which turns out to be:
\begin{equation}
H=H_0\sin[\sin(2mt)]~.
\end{equation}

Though it is difficult to integrate $H$ analytically to get scale
factor $a(t)$, we can already read from the formula above that the
Hubble parameter $H$ is strictly periodical and can cross zero
cyclicly, which means that the universe experiences expanding and
contracting phase alternately. Fig.2 is our numerical results.
Similar to the two-fluid case, we can see that in this case a cyclic
universe is also obtained in the framework of LQC.

\begin{figure}[htbp]
\includegraphics[scale=0.8]{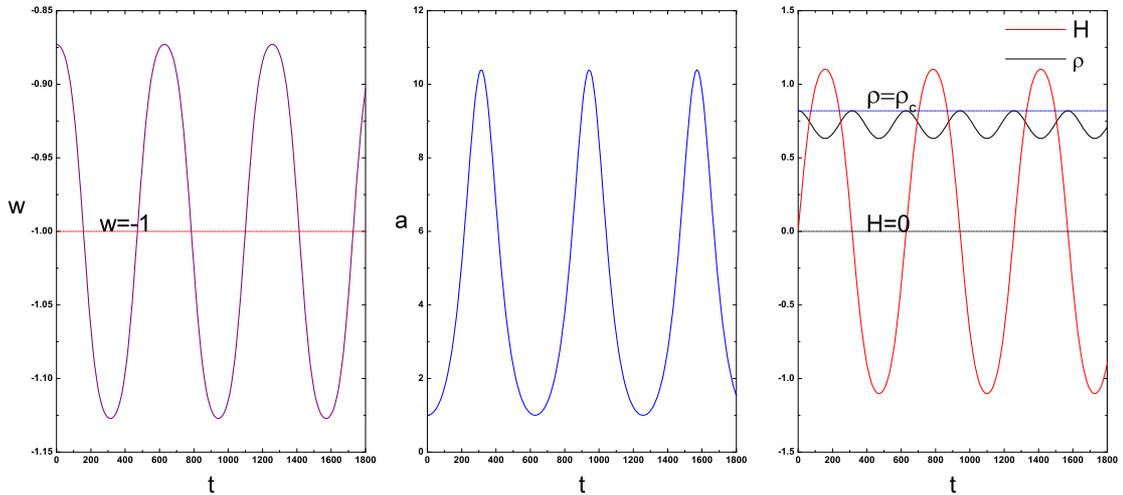}
\caption{Plots of the evolution of the EoS parameter $w$, scale
factor $a$, Hubble parameter $H$ and energy density $\rho$ for the
Quintom model of two scalar fields. In the numerical calculation
we take $m=0.5$, $\lambda=4.80$ and choose the initial values of
parameters to be $\phi_0=0.42$, $\psi_0=0.42$, $V_0=0.41$.}
\label{fig2:eos}
\end{figure}

\section{Discussion and Conclusion}

In this paper, we have studied an application of Quintom matter in
the framework of LQC. Our results show that due to the specific
property of Quintom and with the help of modified Friedmann
equation, a scenario of cyclic universe can be realized naturally.

Before conclusion we should point out that:

1) in the scenario of cyclic universe with Quintom matter in LQC,
there could be two possibilities causing the turnaround or the
bounce, one being $\rho=\rho_c$ as studied in this paper, the
other being $\rho=0$ from Eq.(\ref{b2}). If the energy density
reaches zero during the evolution, the turnaround or bounce can be
realized as well which is pointed out in Ref. \cite{Cai:2007qw}
(also see Ref. \cite{Brown:2004cs}). However, this possibility
will not happen in the examples studied in this paper. For the
double-fluid model, we can see from the formula that the energy
density cannot approach zero; while for the double-field model,
the ``ghost" field $\psi$ has a minus mass squared term, which
prohibits the abnormal kinetic term to be so large to drive the
total energy density to reach zero.

A cyclic universe is a non-standard scenario of cosmology which,
by having the scale factor oscillating and the universe expanding
and contracting alternately, is expected to solve the Big-Bang
singularity and coincidence problem. In the literature there have
been many discussions on such a topic and a number of models have
been proposed, among which there are cyclic models in braneworld
scenario \cite{Steinhardt:2001vw,Piao:2004hr,Piao:2004me}, cyclic
cosmologies with spinor matter \cite{ArmendarizPicon:2003qk},
closed oscillating universe \cite{oscillating-universe,
Clifton:2007tn}, cyclic braneworld in LQC \cite{loop-branecyc,
singh-bounce}, and see Refs. \cite{Zhang:2007an, Saridakis:2007cf}
for recent developments, and so on. The difference of our model
from these models is that our model is in the framework of
4-dimensional spatially flat FRW universe with LQC.

2)  our model provides a qualitative picture of cyclic evolution of
the universe. In this paper we have not included the radiation and
matter, however, the general behavior of our solution will not be
changed since the epochs for radiation and matter only last for a
short time in one period during the whole evolution. There are some
open issues in the scenario of cyclic universe, such as the entropy
problem \cite{Turok:2004yx, Baum:2006nz}, blackhole problem
\cite{Steinhardt:2004gk, Brown:2004cs, Zhang:2007yu}, the problem of
causality \cite{Turok:2004yx} and so on, which is still in
discussion in the literature.

\section*{Acknowledgments}

We thank Robert Brandenberger and Yun-Song Piao for discussions.
This work is supported in part by National Natural Science
Foundation of China under Grant Nos. 90303004, 10533010 and
10675136 and by the Chinese Academy of Science under Grant No.
KJCX3-SYW-N2.

\end{document}